\documentclass{article}
\usepackage{amsmath,amssymb}
\usepackage{graphicx}
\usepackage{textgreek}
\usepackage{authblk}
\usepackage{cite}

\title{On the effect of self-steepening in modulation instability}
\author[1,*]{S. M. Hernandez}
\author[2,3]{P. I. Fierens}
\author[1]{J. Bonetti}
\author[1,3]{D. F. Grosz}
\affil[1]{Grupo de Comunicaciones \'Opticas, Instituto Balseiro,Bariloche, R\'io Negro 8400, Argentina}
\affil[2]{Grupo de Optoelectr\'onica, Instituto Tecnol\'ogico de Buenos Aires,CABA 1106, Argentina}
\affil[3]{Consejo Nacional de Investigaciones Cient\'ificas y T\'ecnicas (CONICET),  Argentina}
\affil[*]{Corresponding author: shernandez@ib.edu.ar}

\begin{document}

\maketitle

% \dates{Compiled \today}

\begin{abstract}
  We revisit the problem of modulation instability (MI) in optical
  fibers, including higher-order dispersion terms, self-steepening,
  and Raman response. We derive expressions for the MI gain and use
  them to explore the role of self-steepening towards a high-power
  limit. We show that, contrary to common wisdom, there is a pump
  power level that maximizes the MI gain. Further increasing the power
  not only diminishes the gain, but eventually makes it disappear. We
  believe these findings to be of special relevance, for instance,
  when applied to the generation of supercontinuum in the mid and far
  infrared bands. Finally, numerical simulations confirming our
  analitycal results are presented.
\end{abstract}

%\begin{document}

%\thispagestyle{fancy}

% \section{Introduction}
% \label{sec:intro}

The phenomenon of modulation instability (MI) has been known and
thoroughly studied for many years in a vast number of different areas
of science. In the realm of optical fibers
\cite{Hasegawa.IEEEJournalQuantumElectronics.1980,Anderson.OpticsLett.1984,Tai.PRL.1986,Potasek.OptLett.1987,Potasek.PRA.1987,Nakazawa.PhysRevA.1989,Agrawal.IEEEPhotonicsTechnologyLett.1992},
MI plays a fundamental role as it is intimately connected to the
appearance of optical solitons, which have had a strong impact on
applications to high-capacity fiber-optics communication. Modulation
instability also is at the heart of the occurrence of efficient
parametric optical processes heavily relied upon to achieve bright and
coherent light in the infrared range. These very same nonlinear
processes are used to provide optical amplification and wavelength
conversion in the telecommunication band, maybe one day enabling
complete photonic control of optical data traffic. In recent years,
nonlinear phenomena such as supercontinuum generation
\cite{Demircan.OpticsComm.2005,Frosz.OptExpress.2006,Frosz.JOSAB.2006,Dudley.OptExpress.2009,Masip.OptLett.2009}
and rogue waves
\cite{Hammani.IEEEPhotonics.2009,Sorensen.JOSAB.2012,Toenger.ScientificReports.2015}
have rekindled the interest in MI.

Despite the abundant literature on the subject, to the
best of our knowledge a complete analysis of MI including both the
Raman response and the effect of self-steepening has only been
presented by B\'ejot et al. \cite{Bejot.PhysRevA.2011}. In this paper,
we derive expressions of the MI gain that coincide with those in
Ref.~\cite{Bejot.PhysRevA.2011} and focus on the role of
self-steepening. By analyzing the dependence of the MI gain with the
input pump power, we find that self-steepening plays a fundamental
role as it yields an optimum power (in terms of maximizing the
gain) and, surprisingly, makes the gain vanish above a certain
threshold, which is obtained from the analytical model.

% \section{Analytical expression of the modulation instability gain}
% \label{sec:migain}

Wave propagation in a lossless optical fiber can be described by the
generalized nonlinear Schr\"{o}dinger equation
\cite{Agrawal.NLFO.2012},
\begin{equation}
  \frac{\partial A}{\partial z}-i\hat{\beta}A
  = i \hat{\gamma} A(z,T)\int\limits_{-\infty}^{+\infty}R(T') \left|A(z,T-T')\right|^2 dT',
  \label{eq:gnlse}
\end{equation}
where $A(z,T)$ is the slowly-varying envelope, $z$ is the spatial
coordinate, and $T$ is the time coordinate in a comoving frame at the
group velocity ($=\beta_1^{-1}$). $\hat{\beta}$ and $\hat{\gamma}$ are
operators related to the dispersion and nonlinearity, respectively,
and are defined by
\begin{equation*}
  \hat{\beta}= \sum_{m\geq 2}\frac{i^m}{m!}\beta_m \frac{\partial^m }{\partial T^m},\; \hat{\gamma} = \sum_{n \geq 0} \frac{i^n}{n!} \gamma_n \frac{\partial^n }{\partial T^n}.
\end{equation*}
The $\beta_m$'s are the coefficients of the Taylor expansion of the
propagation constant $\beta(\omega)$ around a central frequency
$\omega_0$. In the convolution integral in the right hand side of
\eqref{eq:gnlse}, $R(T)$ is the nonlinear response function that
includes both the instantaneous (electronic) and delayed Raman
response.

We shall analyze the effect of a small perturbation $a$ to the
stationary solution $A_s$ of \eqref{eq:gnlse} (see
\cite{Agrawal.NLFO.2012})
\begin{equation}
  A(z,T) = \left(\sqrt{P_0}+a\right) e^{i\gamma_0P_0z} = A_s + a e^{i\gamma_0P_0z}.
  \label{eq:perturbation}
\end{equation}
If we keep only terms linear in the perturbation, after some
manipulations, substitution of \eqref{eq:perturbation} into
\eqref{eq:gnlse} leads to
\begin{equation}
  \frac{\partial \tilde{a}(z,\Omega)}{\partial z} +\tilde{N}(\Omega)\tilde{a}(z,\Omega) = \tilde{M}(\Omega)\tilde{a}^*(z,-\Omega),
  \label{eq:gnlseomegab}
\end{equation}
where $\Omega = \omega - \omega_0$, $\tilde{a}$, $\tilde{\beta}$,
$\tilde{\gamma}$, and $\tilde{R}$ are the Fourier transforms of $a$,
$\beta$, $\gamma$ and $R$, respectively. Moreover, for the sake of
clarity we have defined
\begin{eqnarray*}
  &\tilde{N}(\Omega) &= -i \left[\tilde{\beta}(\Omega)+P_0 \tilde{\gamma}(\Omega) \left(1+\tilde{R}(\Omega)\right)-P_0\gamma_0\right],\\
  &\tilde{M}(\Omega) &= i P_0 \tilde{\gamma}(\Omega)\tilde{R}(\Omega).
  \label{eq:gnlseomega4defS}
\end{eqnarray*}
After some straightforward algebra, \eqref{eq:gnlseomegab} can be cast
into a 2\textsuperscript{nd} order ordinary differential equation
\begin{equation}
  \frac{\partial^2 \tilde{a}(z,\Omega)}{\partial z^2} +2i\tilde{B}(\Omega)\frac{\partial \tilde{a}(z,\Omega)}{\partial z}
  - \tilde{C}(\Omega) \tilde{a}(z,\Omega) = 0,
  \label{eq:gnlseomegaq}
\end{equation}
where
\begin{eqnarray*}
  \tilde{\beta}_e(\Omega) = \sum_{n\geq 1} \frac{\beta_{2n}}{(2n)!} \Omega^{2n},\;\tilde{\beta}_o(\Omega) = \sum_{n\geq 1} \frac{\beta_{2n+1}}{(2n+1)!} \Omega^{2n+1},\\
  \tilde{\gamma}_e(\Omega) = \sum_{n\geq 0} \frac{\gamma_{2n}}{(2n)!} \Omega^{2n},\;\tilde{\gamma}_o(\Omega) = \sum_{n\geq 0} \frac{\gamma_{2n+1}}{(2n+1)!} \Omega^{2n+1},
\end{eqnarray*}
\begin{equation}
  \tilde{B}(\Omega)= -\left[\tilde{\beta}_o(\Omega)+P_0 \tilde{\gamma}_o(\Omega)\left(1+\tilde{R}(\Omega)\right)\right],
  \label{eq:gnlseomegar}
\end{equation}
\begin{equation}
  \begin{split}
    \tilde{C}(\Omega)= &\tilde{\beta}_o^2(\Omega)-\tilde{\beta}_e^2(\Omega)+\\
    +&P_0^2\left(\tilde{\gamma}_o^2(\Omega)-\tilde{\gamma}_e^2(\Omega)\right)\left(1+2\tilde{R}(\Omega)\right)-P_0^2\gamma_0^2+\\
    +&2P_0\gamma_0 \tilde{\beta}_e(\Omega)+2P_0^2\gamma_0 \tilde{\gamma}_e(\Omega)\left(1+\tilde{R}(\Omega)\right)+\\
    +&2P_0\left(\tilde{\beta}_o\tilde{\gamma}_o-\tilde{\beta}_e\tilde{\gamma}_e\right)\left(1+\tilde{R}(\Omega)\right).
  \end{split}
  \label{eq:gnlseomegas}
\end{equation}
Substitution of $a(z,\Omega) = D\exp(iK(\Omega)z)$ in
\eqref{eq:gnlseomegaq} leads to the dispersion relation
\begin{equation}
  K_{1,2}(\Omega) = -\tilde{B}(\Omega)\pm \sqrt{\tilde{B}^2(\Omega)-\tilde{C}(\Omega)}.
  \label{eq:reldispa}
\end{equation}
A simple expression can be obtained by setting $\gamma_n = 0$ for
$n\geq 2$ and $\gamma_1 = \gamma_0 \tau_{\mathrm{sh}}$ (accounting for
the effect of self-steepening). In this case,
\begin{equation}
  \begin{split}
    K_{1,2}(\Omega) = &\tilde{\beta}_o+P_0 \gamma_0 \tau_{\mathrm{sh}}\Omega\left(1+\tilde{R}\right)\pm\\
    \pm&\sqrt{\left(\tilde{\beta}_e+2\gamma_0P_0\tilde{R}\right)\tilde{\beta}_e
      + P_0^2 \gamma_0^2 \tau_{\mathrm{sh}}^2\Omega^2\tilde{R}^2}.
  \end{split}
  \label{eq:reldispb}
\end{equation}
This expression agrees with a similar one presented in
Ref.~\cite{Bejot.PhysRevA.2011} and with the one with
$\tau_{\mathrm{sh}} = 0$ in Ref.~\cite{Frosz.JOSAB.2006}. As usual,
the MI gain can be found as
\begin{equation}
  \label{eq:migaindef}
  g(\Omega) = 2 \max \{-\mathrm{Im}\{K_1(\Omega)\},-\mathrm{Im}\{K_2(\Omega)\},0\},
\end{equation}
where the factor 2 is due to the fact that $g(\Omega)$ is a power gain. The
resulting equation exhibits many properties of the gain that have been
thoroughly studied in the literature, for instance, the fact that it
does not depend on odd terms of the dispersion relation (e.g.,
$\beta_3$) \cite{Potasek.OptLett.1987, Frosz.JOSAB.2006}. However, the
derived MI gain also reveals some novel aspects related to the
self-steepening term $\gamma_0 \tau_{\mathrm{sh}}$. Indeed, it already
has been noted that this term enables a gain even in a zero-dispersion
fiber and that, in general, leads to a narrowing of the MI gain
bandwidth \cite{Abdullaev.OpticsComm.1994,Abdullaev.JOSAB.1997}.

For the case of a large input power, \eqref{eq:reldispb} shows that
the MI gain spectrum is dominated by the Raman response,
\textit{i.e.},
\begin{equation}
  \left|g(\Omega)\right| \approx 4 P_0 \gamma_0 \tau_{\mathrm{sh}}|\Omega|\cdot \left|\mathrm{Im}\left\{\tilde{R}(\Omega)\right\}\right|.
  \label{eq:reldispc}
\end{equation}
Then, in the large pump power limit, the MI gain is independent of the
dispersion parameters $\beta_m$. Although this particular scenario is
worth mentioning, very interesting properties are actually revealed
when studying what happens as we tend towards this large-power
regime. It is widely known \cite{Agrawal.NLFO.2012} that, for the
simplified model that only takes $\beta_2$ and $\gamma_0$ into
account, and no self-steepening, as the pump power $P_0$ increases,
the frequency $\Omega_{\mathrm{max}}$ where the MI gain attains its
maximum and the peak gain, both increase as, respectively,

\begin{equation}
  \label{eq:maxgainsimple}
  \Omega_{\mathrm{max}} = \pm \sqrt{\frac{2 \gamma_0 P_0}{|\beta_2|}},\qquad g(\Omega_{\mathrm{max}}) = 2 \gamma_0 P_0.
\end{equation}
 
Enter self-steepening and the dependence between the pump power and
the MI gain changes drastically in a non-trivial way, for instance,
there is an optimum pump power level for which a peak gain is
attained; any further increase in pump power will make the MI gain
decline. To see this, let us analyze the case of \eqref{eq:reldispb}
considering only the electronic response (\emph{i.e.},
$\tilde{R}(\Omega)=1$). We have
\begin{equation}
  K_{1,2}(\Omega) = \tilde{\beta}_o+2 P_0 \gamma_0 \tau_{\mathrm{sh}}\Omega \pm 
  \sqrt{\Delta},
  \label{eq:reldispsimpl}
\end{equation}
where
\begin{equation}
  \label{eq:Delta}
  \Delta=(\tilde{\beta}_e+2\gamma_0
  P_0)\tilde{\beta}_e + P_0^2 \gamma_0^2
  \tau_{\mathrm{sh}}^2\Omega^2.   
\end{equation}
We have gain whenever the imaginary part of \eqref{eq:reldispsimpl} is
negative, \emph{i.e.}, $\Delta < 0$. It is easily seen that, if we
`turn off' self-steepening, $\Delta = (\tilde{\beta}_e+2\gamma_0
P_0)\tilde{\beta}_e$ and if $\tilde{\beta}_e < 0$ (anomalous
dispersion) there always will exist a sufficiently high $P_0$ such
that $\Delta<0$, namely, $P_0 > \frac{|\tilde{\beta}_e|}{2
  \gamma_0}$. It suffices to find where $\Delta=\Delta(\Omega)$
attains its maxima to find the maximum MI gain. Then,
\begin{equation}
  \label{eq:1}
  \partial_{\Omega} \Delta = 2 \tilde{\beta}_e
  \partial_{\Omega} \tilde{\beta}_e + 2 \gamma_0 P_0 \partial _{\Omega} \tilde{\beta}_e=0.
\end{equation}
$\partial _{\Omega} \tilde{\beta}_e(\Omega)$ need not necessarily be
nonzero for all $\Omega \neq 0$, but we assume so for the frequency
range of interest. In addition, if $\tilde{\beta}_e(\Omega)$ is
negative and decreasing, the sufficient condition for the maximum gain
becomes $\tilde{\beta}_e(\Omega)=-\gamma_0 P_0$ and it is given by
\begin{equation}
  \label{eq:3}
  g(\Omega_{\mathrm{max}}) = 2 \gamma_0 P_0, \mbox{ with } \Omega_{\mathrm{max}}=\{\Omega : \tilde{\beta}_e(\Omega)=-\gamma_0 P_0\},
\end{equation}
which is the same expression as in \eqref{eq:maxgainsimple}, except
that obtaining the desired frequencies is more involved.  Let us now
turn our attention to the effect of self-steepening. Consider
\eqref{eq:Delta}, this time with $\tau_{\mathrm{sh}} \neq 0$. The
necessary condition for gain, namely $\Delta < 0$, becomes
\begin{equation}
  \label{eq:5}
  (\tilde{\beta}_e+2\gamma_0 P_0)\tilde{\beta}_e + P_0^2 \gamma_0^2
  \tau_{\mathrm{sh}}^2\Omega^2 < 0,
\end{equation}
where the rightmost term in the l.h.s. assures that \emph{there will
  be a given pump power level above which gain vanishes}. As in the
case of \eqref{eq:3}, obtaining an explicit analitycal formula for the
MI frequencies is not possible for arbitrary dispersion profiles but,
as before, we can write an implicit formula and obtain the MI peak
gain.  By making $\partial _{\Omega} \Delta=0$, assuming anomalous
dispersion, and $\partial _{\Omega} \tilde{\beta}_e < 0$ in the
frequency range of interest, we obtain
\begin{equation}
  \label{eq:6}
  \Omega_{\mathrm{max}} \in \{\Omega : \tilde{\beta}_e(\Omega) + \gamma_0 P_0 + \Omega \frac{(\gamma_0 P_0 \tau_{\mathrm{sh}})^2} {\partial _{\Omega} \tilde{\beta}_e(\Omega)} = 0 \}.
\end{equation}
Then, by direct substitution, the maximum gain is given by
\begin{equation}
  \label{eq:7}
  g(\Omega_{\mathrm{max}}) = 2 \sqrt{(\gamma_0 P_0)^2 F_{\mathrm{sh}} }
\end{equation}
where
\begin{equation}
  \label{eq:8}
  F_{\mathrm{sh}} = 1 - \Omega_{\mathrm{max}}^2 \tau_{\mathrm{sh}}^2 \left[\frac{(\gamma_0 P_0 \tau_{\mathrm{sh}})^2}{ (\partial _{\Omega} \tilde{\beta}_e(\Omega_{\mathrm{max}}))^2} + 1 \right].
\end{equation}
Even though $F_{\mathrm{sh}}$ is defined implicitly, we can draw some
conclusions.  If $\tau_{\mathrm{sh}} = 0$, then $F_{\mathrm{sh}}=1$
and $g(\Omega_{\mathrm{max}})=2 \gamma_0 P_0$, as expected. Otherwise,
$F_{\mathrm{sh}}$ provides a measure of how the gain is affected by
self-steepening.  Since $F_{\mathrm{sh}} \leq 1$, gain can only be
reduced, for $F_{\mathrm{sh}} \in (0,1)$, or vanish for
$F_{\mathrm{sh}} \leq 0$. As such, $F_{\mathrm{sh}} \geq 0$ is an
alternative condition to \eqref{eq:5} for MI gain to exist.

As an example, let us solve \eqref{eq:8} for the case $\beta_2 < 0,
\beta_{n\geq 3}=0$. An explicit computation can be carried out and
\begin{equation}
  \label{eq:9}
  \Omega_{\mathrm{max}} = \pm \frac{\sqrt{2 |\beta_2| \gamma_0 P_0 - 2(\gamma_0 P_0 \tau_{\mathrm{sh}})^2}}{|\beta_2|}.
\end{equation}
Maxima or, equivalently, gain will occur if and only if $2 |\beta_2|
\gamma_0 P_0 - 2(\gamma_0 P_0 \tau_{\mathrm{sh}})^2 > 0$, i.e.,
\begin{equation}
  \label{eq:10}
  P_0 < \frac{|\beta_2|}{\gamma_0 \tau_{\mathrm{sh}}^2}.
\end{equation}
Any further pump power increase and the medium will exhibit no gain, a
fundamental difference when compared to the case without
self-steepening. In fact, there is an optimal pump power $P_0^o \in
\left(0, |\beta_2|/(\gamma_0 \tau_{\mathrm{sh}}^2)\right)$ for which a
maximum gain is achieved. $P_0^o$ can be obtained by solving $2
F_{\mathrm{sh}} + P_0 \partial_{P_0} F_{\mathrm{sh}} = 0$. For the
case $\tilde{\beta}_e(\Omega)=(\beta_2/2) \Omega^2$,
\begin{equation}
  \label{eq:12}
  P_0^o = \frac{1}{2} \frac{|\beta_2|}{\gamma_0 \tau_{\mathrm{sh}}^2}.
\end{equation}

\begin{figure}[t]
  \includegraphics[width=0.99\linewidth]{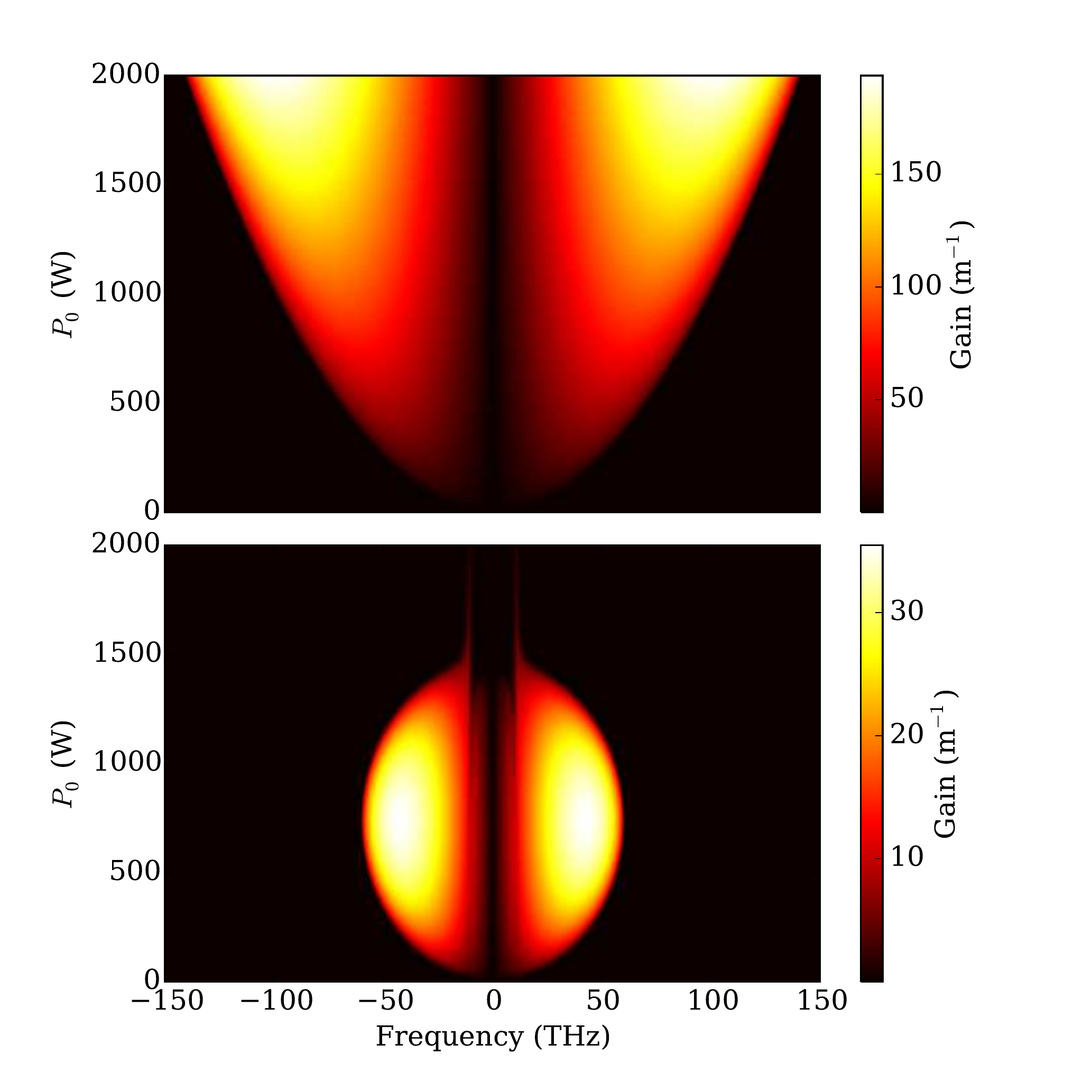}
  \caption{MI gain versus pump power when
    only $\beta_2$ is considered, with (bottom pane) and without (top
    pane) self-steepening.}
  \label{fig:1}
\end{figure}

\begin{figure}[t]
  \includegraphics[width=0.99\linewidth]{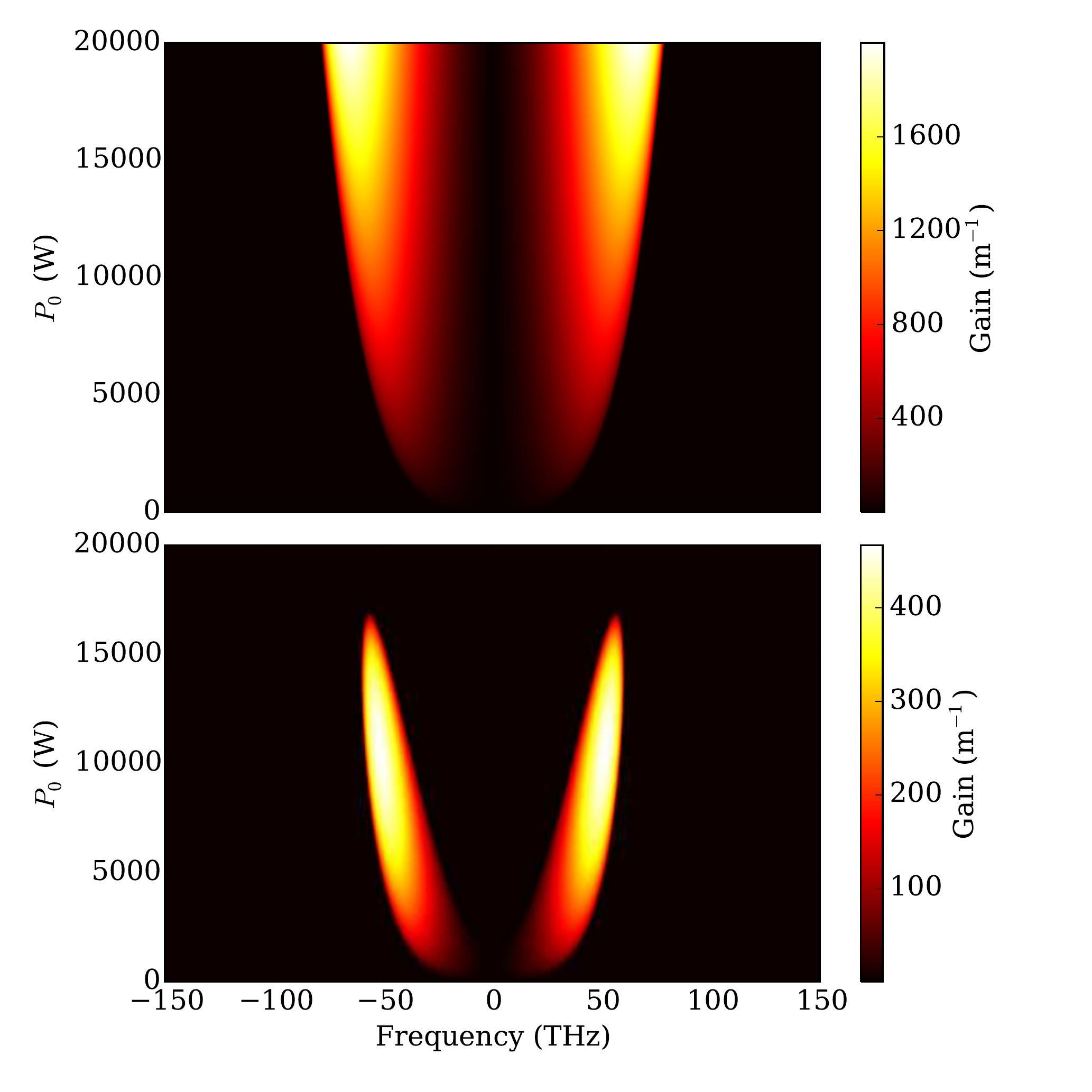}
  \caption{MI gain versus pump power when
    up to $\beta_4$ is considered, with (bottom pane) and
    without (top pane) self-steepening.}
  \label{fig:2}
\end{figure}

\begin{figure}[t]
  \includegraphics[width=0.99\linewidth]{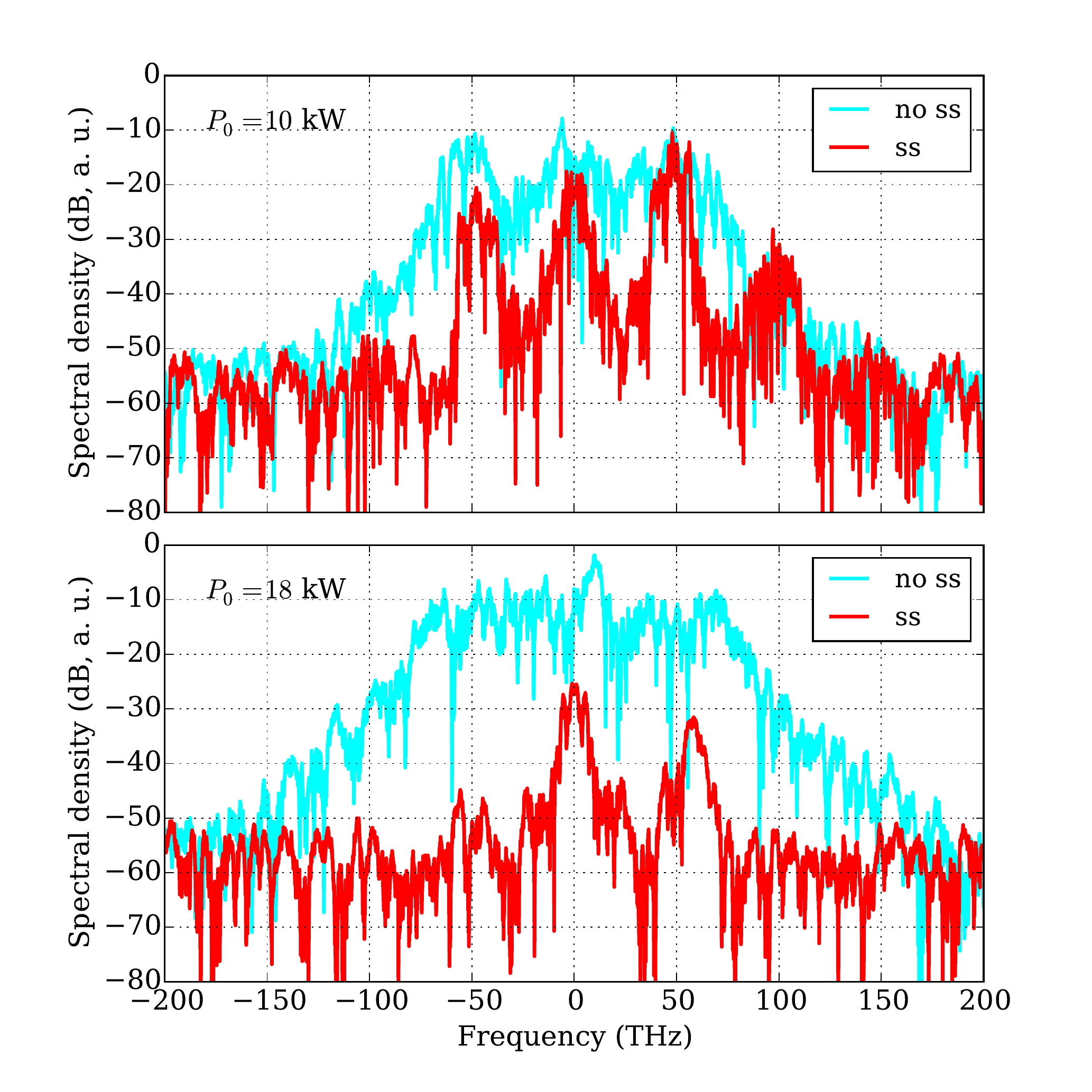}
  \caption{Numerical simulations for the fibers of Fig.~\ref{fig:2}
  at a propagated distance of 10~mm. Pump power of 10 kW (top pane) and
  18 kW (bottom pane) show the effect of (not) considering self-steepening.}
  \label{fig:3}
\end{figure}
\noindent That is, \eqref{eq:12} gives a peak MI gain right in the
middle of the power range for which there is gain.  This can be seen
in Fig.~\ref{fig:1} (bottom pane), where MI gain for a fiber with
$\beta_2 = -1\,\mathrm{ps}^2/\mathrm{km}$, $\beta_{n \geq 3} = 0$,
$\gamma_0=100\,(\mbox{W-km})^{\mathrm{-1}}$, is plotted at different
pump power levels ($P_0$) centered at a wavelength of $5$ \textmu
m. With these parameters \eqref{eq:12} yields $P_0^o \approx
710\;\mathrm{W}$. For comparison, Fig.~\ref{fig:1} also includes the
case without self-steepening (top pane).  We also show the effect of
including a $\beta_4 = -0.0016\,\mathrm{ps}^4/\mathrm{km}$ in
Fig.~\ref{fig:2}. As compared to Fig.~\ref{fig:1}, MI gain bands
appear stretched and the power range changes, but in both cases there
is a power level above which the MI gain vanishes. Both
Figs.~\ref{fig:1}-\ref{fig:2} include the delayed Raman response and
$R(T) = (1-f_R) \delta(T)+f_R h_R(T)$, with
\begin{equation}
  h_R(T) = \frac{\tau_1^2+\tau_2^2}{\tau_1\tau_2^2}e^{-T/\tau_2}\sin(T/\tau_1) u(T),
  \label{eq:raman}
\end{equation} 
where $u(T)$ is the Heaviside step function, $f_R = 0.031$, $\tau_1 =
15.5\,\mathrm{fs}$, $\tau_2 = 230.5\,\mathrm{fs}$
\cite{Karim.OptExpress.2015}. Although
\eqref{eq:reldispsimpl}-\eqref{eq:12} do not include the delayed Raman
response, since the Raman gain spectrum, as given by the Fourier
transform of~\eqref{eq:raman}, provides a much lower and narrower gain
in the lower-frequency limit of the MI gain spectrum, its influence is
much weaker in comparison, as made apparent by the faint Raman gain
bands in Figs.~\ref{fig:1}-\ref{fig:2}.

Figure~\ref{fig:3} shows results of numerical simulations of
\eqref{eq:gnlse} that confirm our observations. Simulation parameters
are those of Fig.~\ref{fig:2}, with $\beta_3 =
0.04\,\mathrm{ps}^3/\mathrm{km}$, at a propagated distance of 10 mm
(for the sake of clarity, the pump was removed from the spectra). When the input pump power is 10 kW (top
pane), with or without considering self-steepening, MI gain is
produced. Nevertheless, when the input pump power level is 18 kW
(bottom pane), and self-steepening is included, MI gain nearly
vanishes. This agrees with the analitycal results depicted in
Fig.~\ref{fig:2} (bottom pane).

% \section{Conclusions}

In summary, we have revisited the problem of modulation instability in
optical fibers. We have included all relevant effects, \textit{i.e.},
higher-order dispersion terms, self-steepening, and the Raman
response. We showed that self-steepening plays a fundamental role in
the large-power limit, and an analytical expression for the pump power
that maximizes the MI gain was derived. Also, we found that, contrary
to common wisdom, increasing the pump power beyond the optimum leads
to a decline in the MI gain and, eventually, to its disappearance. We
believe this observation to be of particular relevance in the case of
supercontinuum generation from CW and quasi-CW sources in the mid and
far infrared, where the effect of self-steepening is expected to be
more pronounced.

% Bibliography
\bibliographystyle{unsrt} \bibliography{biblio}

\end{document}